\documentclass[aps,prl,showpacs,showkeys,amsmath,amssymb,
twocolumn,
floatfix,
superscriptaddress
]{revtex4-1}
\usepackage{graphicx}
\usepackage{times}
\usepackage{verbatim}
\usepackage{color}
\usepackage{cancel}
\usepackage[normalem]{ulem}
\usepackage{bm}
\usepackage{hyperref}
\usepackage{url}

\newcommand{\Eshower}{$E^{\rm rec}_{\mu}$}
\newcommand{\li}{$^{9}$Li}
\newcommand{\he}{$^{8}$He}
\newcommand{\dmee}{$\Delta m^{2}_{ee}$}
\newcommand{\thet}{$\sin^{2}2\theta_{13}$}

\newcommand{\nuebar}{$\overline{\nu}_{e}$~}

\begin{document}

\title{Measurement of electron antineutrino oscillation with 1958 days of operation at Daya Bay}

\newcommand{\ECUST}{\affiliation{Institute of Modern Physics, East China University of Science and Technology, Shanghai}}
\newcommand{\IHEP}{\affiliation{Institute~of~High~Energy~Physics, Beijing}}
\newcommand{\Wisconsin}{\affiliation{University~of~Wisconsin, Madison, Wisconsin 53706}}
\newcommand{\Yale}{\affiliation{Wright~Laboratory and Department~of~Physics, Yale~University, New~Haven, Connecticut 06520}} 
\newcommand{\BNL}{\affiliation{Brookhaven~National~Laboratory, Upton, New York 11973}}
\newcommand{\NTU}{\affiliation{Department of Physics, National~Taiwan~University, Taipei}}
\newcommand{\NUU}{\affiliation{National~United~University, Miao-Li}}
\newcommand{\Dubna}{\affiliation{Joint~Institute~for~Nuclear~Research, Dubna, Moscow~Region}}
\newcommand{\CalTech}{\affiliation{California~Institute~of~Technology, Pasadena, California 91125}}
\newcommand{\CUHK}{\affiliation{Chinese~University~of~Hong~Kong, Hong~Kong}}
\newcommand{\NCTU}{\affiliation{Institute~of~Physics, National~Chiao-Tung~University, Hsinchu}}
\newcommand{\NJU}{\affiliation{Nanjing~University, Nanjing}}
\newcommand{\TsingHua}{\affiliation{Department~of~Engineering~Physics, Tsinghua~University, Beijing}}
\newcommand{\SZU}{\affiliation{Shenzhen~University, Shenzhen}}
\newcommand{\NCEPU}{\affiliation{North~China~Electric~Power~University, Beijing}}
\newcommand{\Siena}{\affiliation{Siena~College, Loudonville, New York  12211}}
\newcommand{\IIT}{\affiliation{Department of Physics, Illinois~Institute~of~Technology, Chicago, Illinois  60616}}
\newcommand{\LBNL}{\affiliation{Lawrence~Berkeley~National~Laboratory, Berkeley, California 94720}}
\newcommand{\UIUC}{\affiliation{Department of Physics, University~of~Illinois~at~Urbana-Champaign, Urbana, Illinois 61801}}
\newcommand{\SJTU}{\affiliation{Department of Physics and Astronomy, Shanghai Jiao Tong University, Shanghai Laboratory for Particle Physics and Cosmology, Shanghai}}
\newcommand{\BNU}{\affiliation{Beijing~Normal~University, Beijing}}
\newcommand{\WM}{\affiliation{College~of~William~and~Mary, Williamsburg, Virginia  23187}}
\newcommand{\Princeton}{\affiliation{Joseph Henry Laboratories, Princeton~University, Princeton, New~Jersey 08544}}
\newcommand{\VirginiaTech}{\affiliation{Center for Neutrino Physics, Virginia~Tech, Blacksburg, Virginia  24061}}
\newcommand{\CIAE}{\affiliation{China~Institute~of~Atomic~Energy, Beijing}}
\newcommand{\SDU}{\affiliation{Shandong~University, Jinan}}
\newcommand{\NanKai}{\affiliation{School of Physics, Nankai~University, Tianjin}}
\newcommand{\UC}{\affiliation{Department of Physics, University~of~Cincinnati, Cincinnati, Ohio 45221}}
\newcommand{\DGUT}{\affiliation{Dongguan~University~of~Technology, Dongguan}}
\newcommand{\XJTU}{\affiliation{Department of Nuclear Science and Technology, School of Energy and Power Engineering, Xi'an Jiaotong University, Xi'an}}
\newcommand{\UCB}{\affiliation{Department of Physics, University~of~California, Berkeley, California  94720}}
\newcommand{\HKU}{\affiliation{Department of Physics, The~University~of~Hong~Kong, Pokfulam, Hong~Kong}}
\newcommand{\UH}{\affiliation{Department of Physics, University~of~Houston, Houston, Texas  77204}}
\newcommand{\Charles}{\affiliation{Charles~University, Faculty~of~Mathematics~and~Physics, Prague}} 
\newcommand{\USTC}{\affiliation{University~of~Science~and~Technology~of~China, Hefei}}
\newcommand{\TempleUniversity}{\affiliation{Department~of~Physics, College~of~Science~and~Technology, Temple~University, Philadelphia, Pennsylvania  19122}}
\newcommand{\CUC}{\affiliation{Instituto de F\'isica, Pontificia Universidad Cat\'olica de Chile, Santiago}} 
\newcommand{\CGNPG}{\affiliation{China General Nuclear Power Group, Shenzhen}}
\newcommand{\NUDT}{\affiliation{College of Electronic Science and Engineering, National University of Defense Technology, Changsha}} 
\newcommand{\IowaState}{\affiliation{Iowa~State~University, Ames, Iowa  50011}}
\newcommand{\ZSU}{\affiliation{Sun Yat-Sen (Zhongshan) University, Guangzhou}}
\newcommand{\CQU}{\affiliation{Chongqing University, Chongqing}} 
\newcommand{\BCC}{\altaffiliation[Now at ]{Department of Chemistry and Chemical Technology, Bronx Community College, Bronx, New York  10453}} 
\author{D.~Adey}\IHEP
\author{F.~P.~An}\ECUST
\author{A.~B.~Balantekin}\Wisconsin
\author{H.~R.~Band}\Yale
\author{M.~Bishai}\BNL
\author{S.~Blyth}\NTU\NUU
\author{D.~Cao}\NJU
\author{G.~F.~Cao}\IHEP
\author{J.~Cao}\IHEP
\author{Y.~L.~Chan}\CUHK
% ...

\collaboration{The Daya Bay Collaboration}\noaffiliation
\date{\today}

\begin{abstract}
\noindent We report a measurement of electron antineutrino oscillation from the Daya Bay Reactor Neutrino
Experiment with nearly 4 million reactor \nuebar inverse $\beta$~decay candidates observed over 1958 days of data collection.
The installation of a Flash-ADC readout system and a special calibration campaign using
different source enclosures reduce uncertainties in the absolute energy calibration
to less than 0.5\% for visible energies larger than 2~MeV\@.
The uncertainty in the cosmogenic $^9$Li and $^8$He background
is reduced from 45\% to 30\% in the near detectors.
A detailed investigation of the spent nuclear fuel history improves its uncertainty from 100\% to 30\%.
Analysis of the relative \nuebar rates and energy spectra among detectors yields
 \thet $= 0.0856\pm 0.0029$ and $\Delta
m^2_{32}=(2.471^{+0.068}_{-0.070})\times10^{-3}~\mathrm{eV}^2$ assuming the normal hierarchy, and
$\Delta m^2_{32}=-(2.575^{+0.068}_{-0.070})\times10^{-3}~\mathrm{eV}^2$ assuming the inverted hierarchy.

\end{abstract}

\maketitle

Neutrino flavor oscillation driven by the $\theta_{13}$ mixing angle has
been observed using reactor antineutrinos~\cite{DB, RENO, Abe:2011fz}
and accelerator neutrinos~\cite{Abe:2013hdq,Adamson:2013ue}. The Daya
Bay experiment previously reported the observation
of a nonzero value of \thet~via the disappearance of reactor
antineutrinos over $\sim$kilometer distances~\cite{DB}, and
a measurement of the effective mass-squared difference
\dmee~via the distortion of the \nuebar energy
spectrum~\cite{DBPRL2014}.
Both of these measurements based on 1230 days of operation are described in detail in Ref.~\cite{Dayabay:2016prd}.
This Letter presents a measurement of these two parameters with a data set
acquired in 1958 days of stable operation,
and with several improvements to the analysis when compared with previous measurements.

The Daya Bay experiment consists of eight
identically designed antineutrino detectors~(ADs),
two in each near experimental hall~(EH1 and EH2), and four in the far hall (EH3).
Antineutrinos are produced by six reactor cores, with two cores $\sim$365~m from EH1, and
four cores $\sim$505~m from EH2. The average geometric baseline to EH3 over all six cores is 1663~m.
Each AD consists of three nested
cylindrical vessels. The inner acrylic vessel is filled with 20-t 0.1\%
gadolinium-doped liquid scintillator~(Gd LS), which serves as the
primary \nuebar target. The acrylic vessel surrounding the target is
filled with undoped LS, increasing the efficiency for detecting $\gamma$
rays produced in the target. A total of 192 8-in.~photomultiplier tubes~(PMTs) are
uniformly positioned on the cylinder of the outermost stainless steel vessel and immersed in mineral oil.
The experimental setup is described in detail in Refs.~\cite{dyb:det,Dayabay:2014vka}.


Stable data taking began on 24 December 2011 with six ADs.
The final two ADs were installed between 28 July and 19 October 2012
in EH2 and EH3. Operation of the 8-AD configuration continued until
20 December 2016. A special calibration was performed
to better understand the neutron detection efficiency~\cite{DYBFlux2018}
and the optical shadowing of the calibration source enclosures.
In January 2017, the Gd LS in EH1 AD1 was replaced by purified LS to
study the light yields with different recipes and purification methods
for next generation experiments.
The remaining seven ADs resumed data taking on 26 January 2017.
Since the statistical precision of the oscillation measurement is
driven by the interaction rate in the far detectors, the impact
of removing EH1-AD1 is negligible. The results presented in this Letter combine
the data from all three periods:
6 AD~(217 days), 8 AD~(1524 days), and 7 AD~(217 days).

Reactor antineutrinos are detected via the inverse $\beta$-decay (IBD)
reaction, $\overline{\nu}_{e} + p \to e^{+} + n$. The positron deposits
kinetic energy in the scintillator and annihilates with an electron, forming a prompt signal,
which gives an estimate of the \nuebar energy,
E$_{\overline{\nu}_e} \approx$ E$_{\rm prompt}$ + 0.78~MeV\@.
Neutron capture on Gd emits several $\gamma$ rays with
a mean total energy of 8.05~MeV, forming
a delayed signal with a mean capture time of $\sim$30~$\mu$s.
The coincidence between the prompt and delayed signals efficiently discriminates
IBD reactions from background.


The primary goal of energy calibration is
to reduce potential energy scale variations among the ADs, which dominate
the systematic error in the oscillation measurement.
PMT gains are calibrated once per day using thermal noise hits.
Light yields ($\sim$160~photoelectrons/MeV) are found to be decreasing by $\sim$1\% per year,
and are calibrated with gamma rays from $^{60}$Co sources deployed weekly.
The 15\% spatial nonuniformity, and its $<$0.5\% per year time-dependence,
are corrected with $^{60}$Co sources along three calibration axes.
Calibration references with different spatial distributions are examined,
including Gd and $^1$H neutron captures, as well as artificially and naturally occurring $\alpha$ and $\gamma$ particles.
The reconstructed energies for these particles are compared among
ADs and found to be consistent within 0.2\%,
which is taken as the energy scale uncertainty uncorrelated among detectors.
Consistent results are obtained by an alternative method using spallation neutron capture on Gd
to calibrate the energy scale, time dependence, and nonuniformity\cite{Dayabay:2016prd}.

The absolute energy response, which relates the actual and observed prompt energies, is improved.
The readout system underestimates the charge of the PMT signals
when they overlap in time due to AC coupling of the front end electronics.
This results in a nonlinear response of the charge over the entire detector at the $\sim$10\% level in the
energy region of interest. Effects of scintillator quenching and Cherenkov radiation contribute an additional $\sim$10\% nonlinearity.
The energy nonlinearity model with 1\% precision used
in previous results is described in Ref.~\cite{Dayabay:2016prd}.
The model is constructed with energies of $\gamma$-rays from deployed and natural
sources, and the $\beta$~spectrum from cosmogenic
$^{12}$B. To improve the understanding of both electronics and LS nonlinearity, dedicated calibrations have been performed.

In December 2015, a full Flash-ADC (FADC) readout system was installed
in EH1 AD1, recording PMT waveforms at 1 GHz and 10-bit resolution.
The FADC and the existing electronics readout system acquire data simultaneously.
A deconvolution method is applied to the waveforms to minimize the dependence on the
single photoelectron pulse shape, in particular the overshoot,
and to extract the integrated charge with minimum bias~\cite{DYBFADC}.
The residual nonlinearity in the reconstructed charge from a single waveform
is estimated to be less than 1\% from an electronics simulation tuned to data.
An event-by-event comparison of the total charge of the two readout systems gives
a measurement of the existing system's nonlinearity at 0.2\% precision.

Uncertainties in the visible energy from $\gamma$~rays~(previously~$\sim$1\%) are dominated by the poor knowledge
of optical shadowing by source enclosures (5~cm tall and 2~cm in diameter cylinders).
A special calibration campaign in January 2017 deployed $^{60}$Co sources
with PTFE, greenish Teflon, and stainless steel enclosures
absorbing $<$0.10\%, 1.22\%, and 0.65\% of photons respectively as determined from simulation.
The reconstructed energies of data and simulation
agree to 0.2\% for all source enclosures.
The total uncertainty from these $\gamma$~rays is improved to 0.5\%, including the residual
nonuniformity between point-like $\gamma$-ray sources, which preferentially illuminate the detector center, 
and IBD events over the full target volume.

The $\beta$ decay of $^{12}$B is an allowed transition of the Gamow-Teller type
with a $Q$~value of 13.4~MeV\@.
A total of 470~000 cosmogenic $^{12}$B candidates
are observed in the Gd LS of the four near ADs. The $\beta$~spectrum of $^{12}$B decay is compared
to a prediction that includes Fermi motion, screening effects, corrections for the finite
size of the nucleus, and weak magnetism.
A significant uncertainty in the prediction is due to the weak magnetism correction,
estimated as a linear correction with a coefficient of (0.48$\pm$0.24)\%/MeV~\cite{Huber:2011,Mueller:2011}.
This uncertainty is propagated to the nonlinearity model together with uncertainties in the decay branching ratios.

\begin{figure}
  \includegraphics[width=0.9\columnwidth]{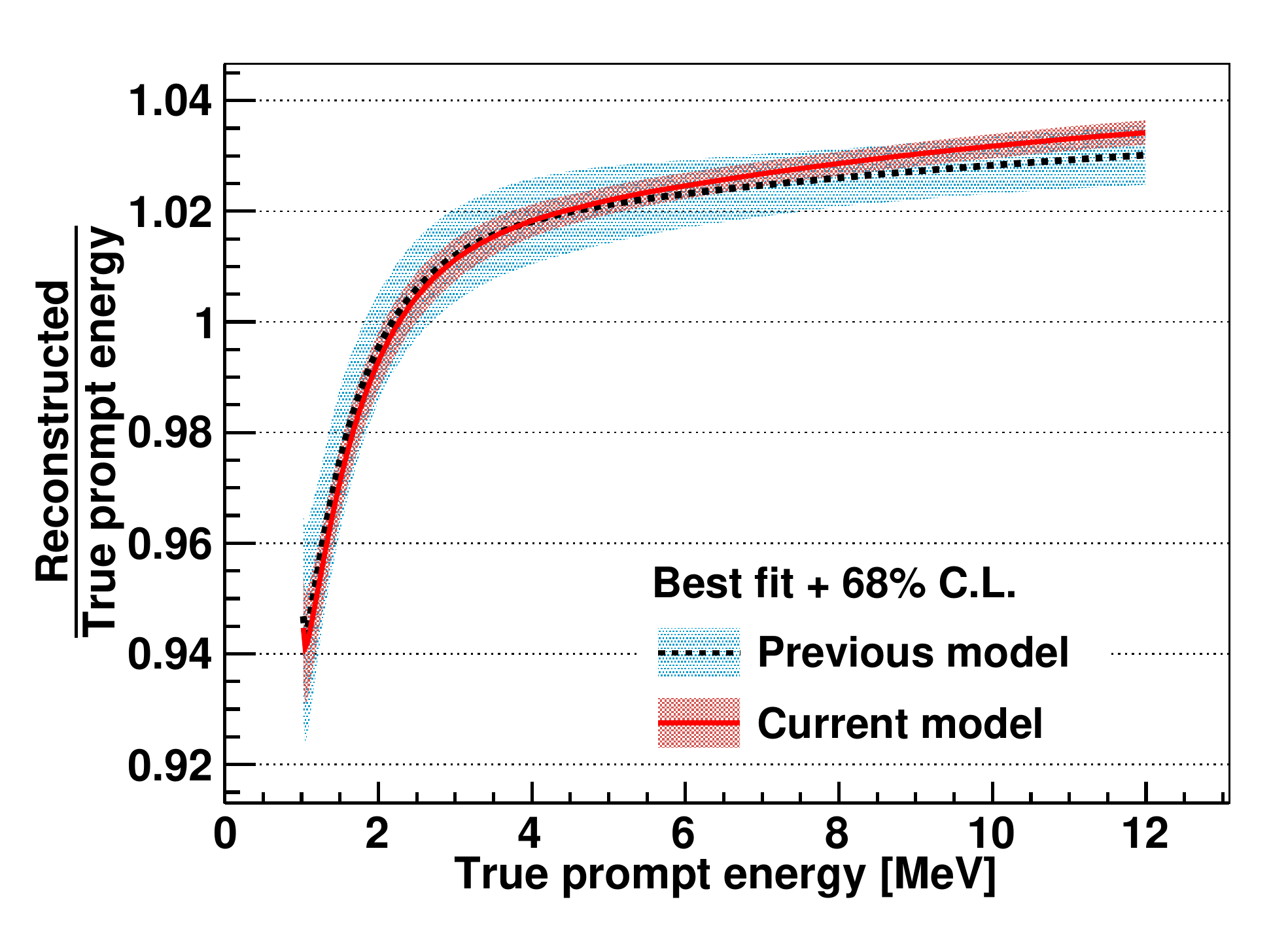}
  \caption{Relationship between the reconstructed and true prompt energy, which is a combination of positron
			kinetic energy and energies of the annihilation $\gamma$~rays. The updated model and its uncertainty~(red) contains improvements described in the text. The previous model~\cite{Dayabay:2016prd}~is shown for comparison.
		 \label{fig:ibdresponse}}
\end{figure}

The functional form of the nonlinearity model used in this analysis is
identical to the one reported in Ref.~\cite{Dayabay:2016prd}.
The improved prompt energy response is shown in Fig.~\ref{fig:ibdresponse}.
The precision is better than 0.5\% for prompt energy larger than 2~MeV\@.
In-flight annihilation and the 3~$\gamma$-rays decay from ortho-positronium
have a $<$0.1\% impact on the energy model. The precision is limited by the systematic uncertainties associated with the $\gamma$-ray samples at prompt energies below 3~MeV, and by the statistics of the $^{12}$B sample at higher energy.
Consistent results are obtained with the removal of any one $\gamma$~ray,
the measured electronics linearity, or the $^{12}$B constraint. Tabulated form of the model is provided in the Supplemental Material~\cite{Supple}.

IBD candidates are selected following the same criteria as
Selection A in Ref.~\cite{Dayabay:2016prd}.
The estimated signal and background rates, as well as the
efficiencies of the muon veto, $\epsilon_{\mu}$, and multiplicity
selection, $\epsilon_m$, are summarized in Table~\ref{tab:ibd}. More than
3.9$\times$10$^{6}$ \nuebar candidates are identified. In all three halls, the background is smaller than 2\%,
and contributes less than 0.15\% to the uncertainty on the IBD rate.
Consistent results are obtained using Selection B in Ref.~\cite{Dayabay:2016prd}.

\begin{figure}
  \includegraphics[width=0.9\columnwidth]{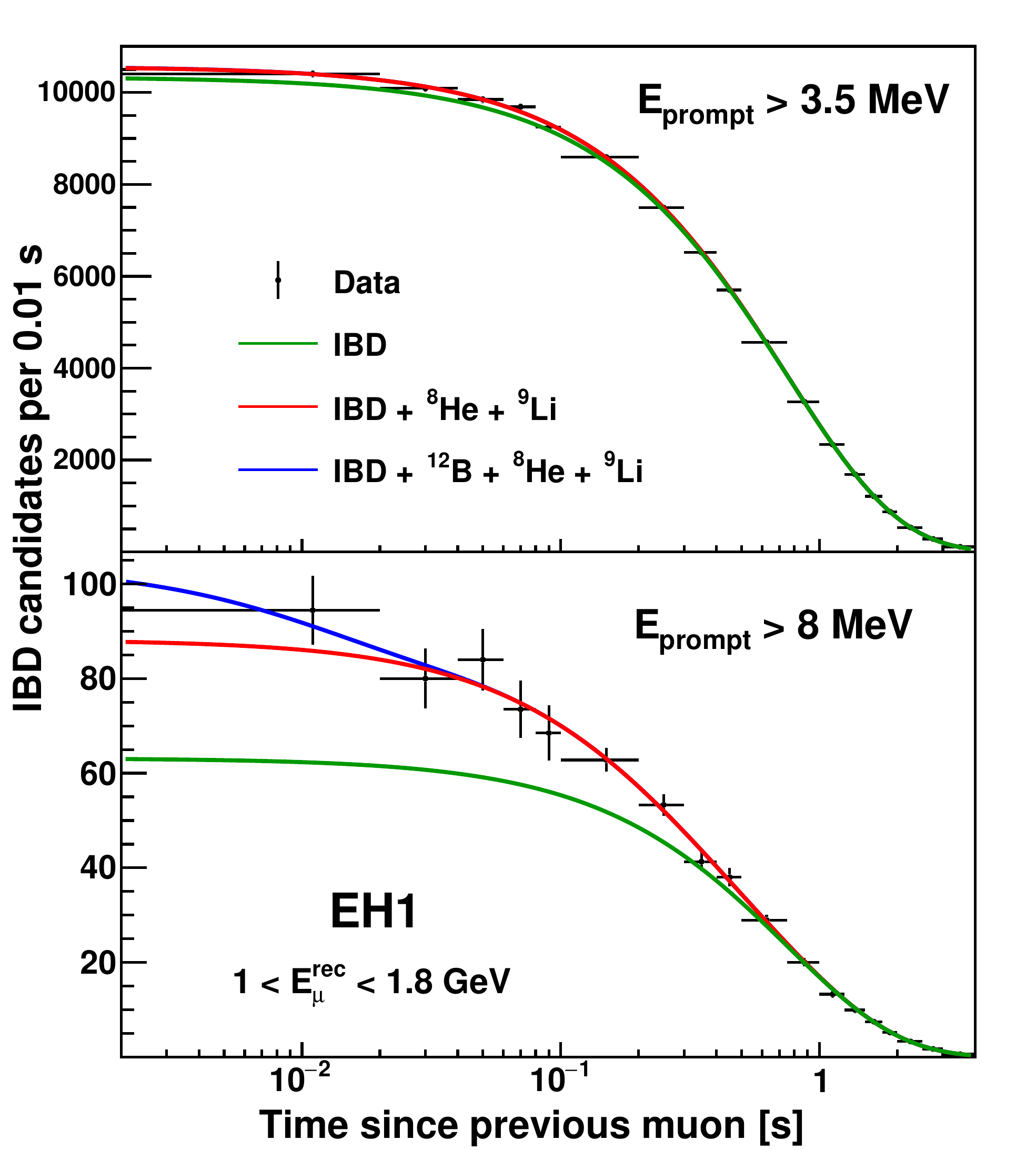}
  \caption{ Time from the IBD candidate to the preceding muon with 1~$<$\Eshower$<$1.8~GeV in EH1.
      Due to the high muon rate and high IBD rate, the \li~component (red curve) cannot be accurately measured
      with a prompt energy cut of 3.5~MeV~(top panel). Applying a higher prompt energy cut of 8~MeV
      reduces the IBD component (green curve), and $^{9}$Li yields can be directly determined~(bottom panel).
      The $^{12}$B~component is due to a coincidence of two $^{12}$B decays produced in the muon shower~(blue curve).
           \label{fig:Li9}}
\end{figure}

The dominant background uncertainty is due to
the cosmogenic production of \li~and \he~(referred as \li~in the following) with subsequent
$\beta$-n decay, which cannot be distinguished from IBD on an event-by-event basis.
Yields are estimated by fitting the distribution of time between the IBD candidate
and the preceding muon, as shown in Fig.~\ref{fig:Li9} for muons with visible energy \Eshower~
between 1 and 1.8~GeV in EH1\@. The falloff with increasing
time depends only on the muon rate for muon-uncorrelated events, while muon-induced \li~
decays with a lifetime of 257~ms.

A prompt energy cut is applied to enhance
the contribution from \li, which has a higher energy spectrum. In the previous analysis,
the cut was 3.5 to 12~MeV, resulting in the distribution shown in the top panel of Fig.~\ref{fig:Li9}.
Due to the low rate of \li~compared to IBD,
it is not possible to determine the \li~rate directly. Instead, an additional neutron
capture signal was required to reduce the muon rate. However, the efficiency of this
additional requirement for true \li~is uncertain, and was the dominant uncertainty in the
previous measurement at 40\%. In this updated analysis with additional statistics,
the prompt energy cut is increased to 8~MeV (6~MeV) to 12~MeV in the near halls (far hall).
This cut suppresses IBD while preserving ($15 \pm 2$)\% of \li~in the near halls and
($40 \pm 3$)\% in the far hall. These efficiencies are determined from the \li~prompt energy 
spectrum, which is measured in data by selecting a \li-enhanced sample immediately following high-energy 
muon showers, and subtracting an IBD-pure spectrum from candidates $>$ 1~s from a muon shower.
The uncertainties on the prompt cut efficiencies are dominated by the statistics of the \li-enhanced sample.
The bottom panel of Fig.~\ref{fig:Li9} shows the high prompt energy cut applied in EH1.
With the higher prompt energy cut, the \li~rate can be determined without requiring a neutron tag for \Eshower $>$ 1~GeV.

In the final \nuebar sample, a veto is applied for 1~ms following muons with 0.02 $<$ \Eshower $<$ 2.5 GeV, and
1~s for muons with \Eshower $>$ 2.5~GeV. With these vetoes, 16\% of the \li~
background arises from muons with \Eshower $<$ 1~GeV, and is estimated using the neutron
tag strategy of Ref.~\cite{Dayabay:2016prd} with a (95$\pm$40)\% efficiency. The improved \li~estimate in
Table~\ref{tab:ibd} is consistent with the previous analysis~\cite{Dayabay:2016prd},
and the uncertainty is reduced from 43\% to 27\% in the near ADs.
The updated uncertainty is dominated by the statistics of the \li-enhanced sample,
and can be further reduced by including additional data.
The largest systematic uncertainty for \Eshower $>$ 1~GeV is 13\%, due to the efficiency of the prompt energy cut.

The detection efficiencies are described in detail in Ref.~\cite{Dayabay:2016prd}.
Correlated uncertainties in efficiency
between the near and far sites cancel in the oscillation measurement.
The 0.2\% uncertainty in the relative energy scale introduces
an uncorrelated 0.08\% uncertainty in the efficiency of
the delayed energy cut. The uncertainty in the Gd capture fraction
is estimated to be 0.1\% by comparing the capture time distributions of the ADs.
Total variation in efficiency among detectors is estimated to be 0.13\%, as in Ref.~\cite{Dayabay:2016prd}.
The measured \nuebar rates are compared between adjacent detectors in the near experimental
halls, in which the statistical uncertainty is of the same order as the uncertainty in the efficiency.
The measured ratios are found to be 0.981$\pm$0.002 between AD1 and AD2~(in the 6- and 8-AD periods)~and 1.014$\pm$0.002
between AD3 and AD8~(in the 7 and 8 AD periods). The expected ratios are 0.982 and 1.013, respectively, accounting for small differences in
the baselines and target mass. Comparison in the far hall is also consistent between data and prediction, but dominated by statistical uncertainty.
The consistency between measured and predicted ratios provides an independent confirmation of
the estimation of the uncorrelated efficiency uncertainty.

\begin{table*}[!htb]
\caption{Summary of signal and backgrounds. Rates are corrected for the muon veto and multiplicity selection efficiencies $\varepsilon_{\mu}\cdot\varepsilon_{m}$. The procedure for estimating accidental, fast neutron, Am-C, and ($\alpha$,n) backgrounds is unchanged from Ref.~\cite{Dayabay:2016prd}.
\label{tab:ibd}}
  \begin{minipage}[c]{\textwidth}
  \resizebox{\textwidth}{!}{
\begin{tabular}{c|cc|cc|cccc} \hline
\hline
  & \multicolumn{2}{c|}{EH1}&\multicolumn{2}{c|}{EH2}&\multicolumn{4}{c}{EH3} \\
  & AD1  & AD2  & AD3 & AD8 & AD4 & AD5 & AD6 & AD7 \\
\hline
\nuebar candidates & 830036	&964381	&889171&784736&127107	&127726	&126666&113922 \\
DAQ live time (days) & 1536.621 &1737.616 &1741.235 &1554.044 &1739.611 &1739.611 &1739.611 &1551.945  \\
$\varepsilon_{\mu}\times \varepsilon_{m}$ & 0.8050&	0.8013&	0.8369&0.8360&0.9596&0.9595&0.9592&0.9595\\
Accidentals (day$^{-1}$) & $8.27\pm0.08$ & $8.12\pm0.08$ & $6.00\pm0.06$ & $5.86\pm0.06$ & $1.06\pm0.01$ & $1.00\pm0.01$ & $1.03\pm0.01$ & $0.86\pm0.01$ \\
Fast neutron (AD$^{-1}$ day$^{-1}$) & \multicolumn{2}{c|}{$0.79\pm0.10$} & \multicolumn{2}{c|}{$0.57\pm0.07$} & \multicolumn{4}{c}{$0.05\pm0.01$} \\
$^9$Li/$^8$He (AD$^{-1}$ day$^{-1}$) & \multicolumn{2}{c|}{$2.38\pm0.66$} & \multicolumn{2}{c|}{$1.59\pm0.49$} & \multicolumn{4}{c}{$0.19\pm0.08$} \\
Am-C correlated(day$^{-1}$) & $0.17\pm0.07$ & $0.15\pm0.07$ & $0.14\pm0.06$ & $0.13\pm0.06$ & $0.06\pm0.03$ & $0.05\pm0.02$ & $0.05\pm0.02$ & $0.04\pm0.02$ \\
$^{13}$C($\alpha$, n)$^{16}$O (day$^{-1}$) & $0.08\pm0.04$ & $0.06\pm0.03$ & $0.04\pm0.02$ & $0.06\pm0.03$ & $0.04\pm0.02$ & $0.04\pm0.02$ & $0.04\pm0.02$ & $0.04\pm0.02$ \\ \hline
\nuebar rate (day$^{-1}$) & $659.36\pm1.00$ & $681.09\pm0.98$ & $601.83\pm0.82$	& $595.82\pm0.85$ & $74.75\pm0.23$ & $75.19\pm0.23$ & $74.56\pm0.23$	& $75.33\pm0.24$ \\
\hline
\end{tabular}}
  \end{minipage}
\end{table*}


The predicted \nuebar flux includes a $\sim$0.3\% contribution from the spent
nuclear fuel present in the cooling pool adjacent to each reactor core~\cite{DYBSpec2017}.
In Ref.~\cite{Dayabay:2016prd}, the uncertainty of the spent fuel contribution was conservatively set to 100\% due to
a lack of knowledge of the spent fuel inventory history.
An investigation of the history, in collaboration with
the nuclear power plant, results in a reduced uncertainty of 30\%,
now dominated by the calculation of the \nuebar spectrum~\cite{DYBSNF}, which sums the
beta decay spectra of fission isotopes with half-lives longer than 10~hours.

Oscillation parameters are extracted from the disappearance of
$\overline{\nu}_{e}$, as given by the survival probability

\begin{align}
P &= 1 - \cos^4\theta_{13}\sin^2 2\theta_{12} \sin^2\Delta_{21} \\
  &~~~~- \sin^2 2\theta_{13}(\cos^2 \theta_{12} \sin^2 \Delta_{31} + \sin^2 \theta_{12} \sin^2\Delta_{32}) \nonumber \\
  &\simeq 1 - \cos^4\theta_{13}\sin^2 2\theta_{12} \sin^2\Delta_{21} \nonumber \\
  &~~~~- \sin^2 2\theta_{13}\sin^2\Delta_{ee}, \nonumber
\end{align}

\noindent
where $\Delta_{ij} = 1.267 \Delta m^{2}_{ij} L/E$. Here
$E$ is the \nuebar energy in MeV and $L$ is the distance in
meters from the \nuebar production point. The oscillation phases due to
$\Delta m^{2}_{31}$ and $\Delta m^{2}_{32}$ are degenerate in the range of $L/E$ relevant for
this measurement. An effective neutrino mass-squared difference
\dmee~in the supplemental material in Ref.~\cite{DBPRL2015}.
Solar parameters $\theta_{12}$~and $\Delta m^2_{21}$~are fixed to the global average in Ref.~\cite{PDG2018},
and their uncertainties are negligible to this measurement.


The observed background-subtracted signal $N_{i}^{far,obs}$ in the $i_{th}$ energy bin in the far hall
is compared to the prediction $N_{i}^{far,pred}$, given in Eq.~\ref{eq:prediction}:

\begin{equation}
  \label{eq:prediction}
  N^{\mathrm{far,pred}}_i = w_{i}\left(\theta_{13},{\Delta}m^2_{\mathrm{ee}}\right) \times N^{\mathrm{near,obs}}_i.
\end{equation}

\noindent
The predicted rate is
based on the measurements in the near halls, $N_{i}^{near,obs}$, with minimal dependence on models of the
reactor \nuebar flux. Weight factors $w_i$ account for the difference in near and far hall measurements, including
detection efficiencies, target mass differences, reactor power and distance from each core, and oscillation probability.
The 6, 8, and 7 AD periods are treated separately in order to properly handle correlations in reactor \nuebar flux,
detector response, and background.

To evaluate the oscillation parameters, a $\chi^2$ is defined in Eq.~\ref{eq:chi2},
where the statistical component of the covariance matrix $V$ is
estimated analytically, and the systematic component is evaluated from simulations:

\begin{equation}
  \label{eq:chi2}
  \chi^{2} = \sum_{i,j}(N_j^{\mathrm{far, obs}} - N_j^{\mathrm{far, pred}})
  (V^{-1})_{ij} (N_i^{\mathrm{far, obs}} - N_i^{\mathrm{far, pred}}).
\end{equation}

\noindent
This approach is described in detail as Method A in Ref.~\cite{Dayabay:2016prd}.

\begin{figure}[!htb]
\centering
\includegraphics[width=0.9\columnwidth]{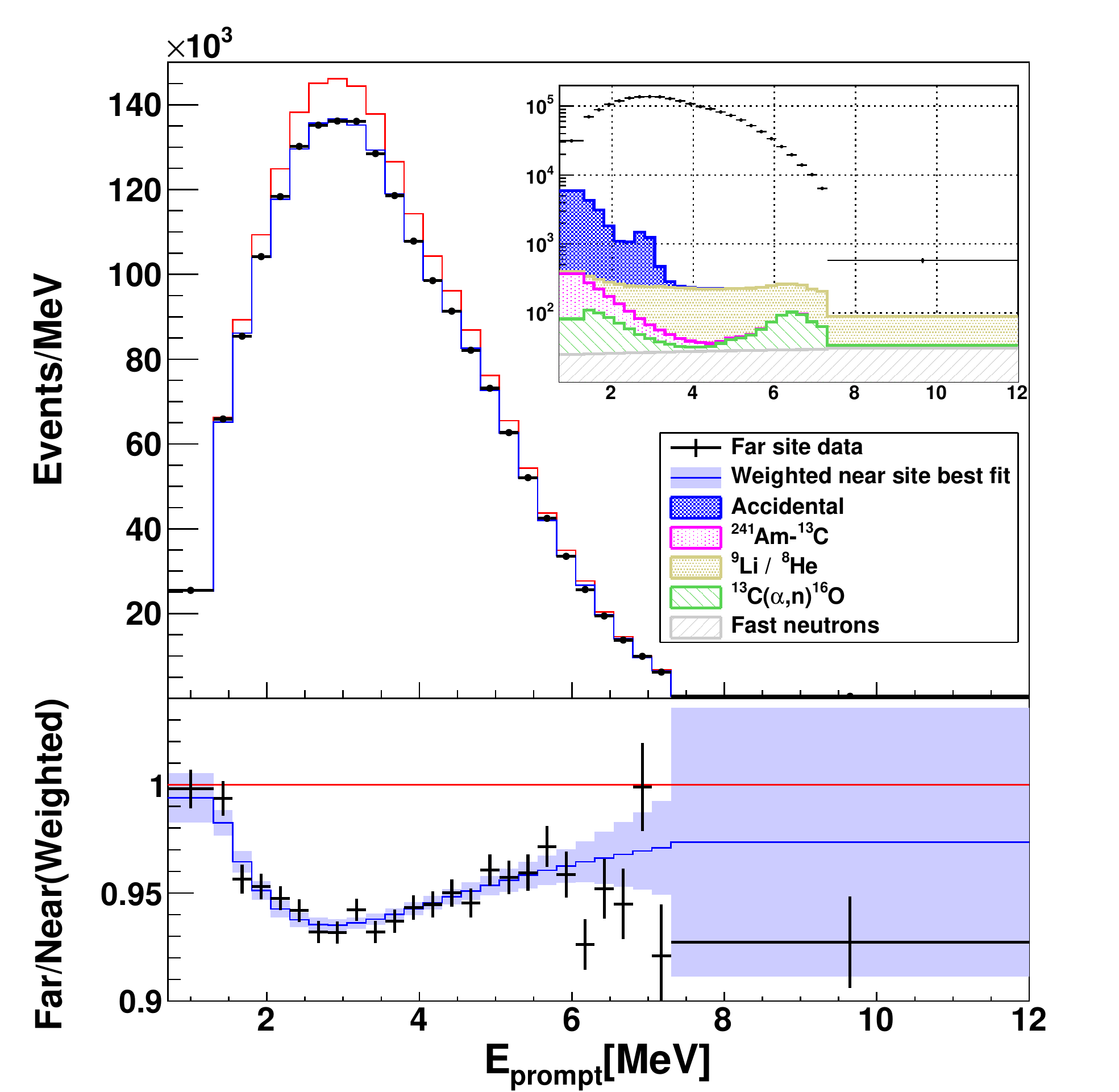}
\caption{The background-subtracted spectrum at the far site~(black points)
         and the expectation derived from near-site measurements excluding~(red line) or
         including~(blue line) the best-fit oscillation. The bottom panel shows
        the ratios of data over predictions with no oscillation. The shaded area
		is the total uncertainty from near-site measurements and the extrapolation model.
		The error bars represent the statistical uncertainty of the far-site data.
        The inset shows the background components on a logarithmic scale. Detailed
		spectra data are provided as Supplemental Material~\cite{Supple}.}
\label{fig:spectracomp}
\end{figure}

Using this method, values of \thet=0.0856$\pm$0.0029 and
\dmee=(2.522$^{+0.068}_{-0.070}$)$\times$10$^{-3}$~eV$^2$ are obtained,
with $\chi^2/\mathrm{NDF}=148.0/154$.
Consistent results are obtained using Methods B or C in Ref.~\cite{Dayabay:2016prd}.
Analysis using the exact \nuebar disappearance probability for three-flavor oscillations yields
$\Delta m^2_{32}=(2.471^{+0.068}_{-0.070})\times10^{-3}~\mathrm{eV}^2$~
($\Delta m^2_{32}=-(2.575^{+0.068}_{-0.070})\times10^{-3}~\mathrm{eV}^2$)
assuming normal (inverted) hierarchy.
Statistics contribute 60\% (50\%) to the total uncertainty in the
\thet~(\dmee) measurement.
The systematic uncertainty of \thet~is dominated by the detection efficiency
uncertainty uncorrelated among detectors and the reactor \nuebar flux prediction, while
that of \dmee~is dominated by the uncorrelated energy scale uncertainty.

\begin{figure}[!htb]
\centering
\includegraphics[width=0.9\columnwidth]{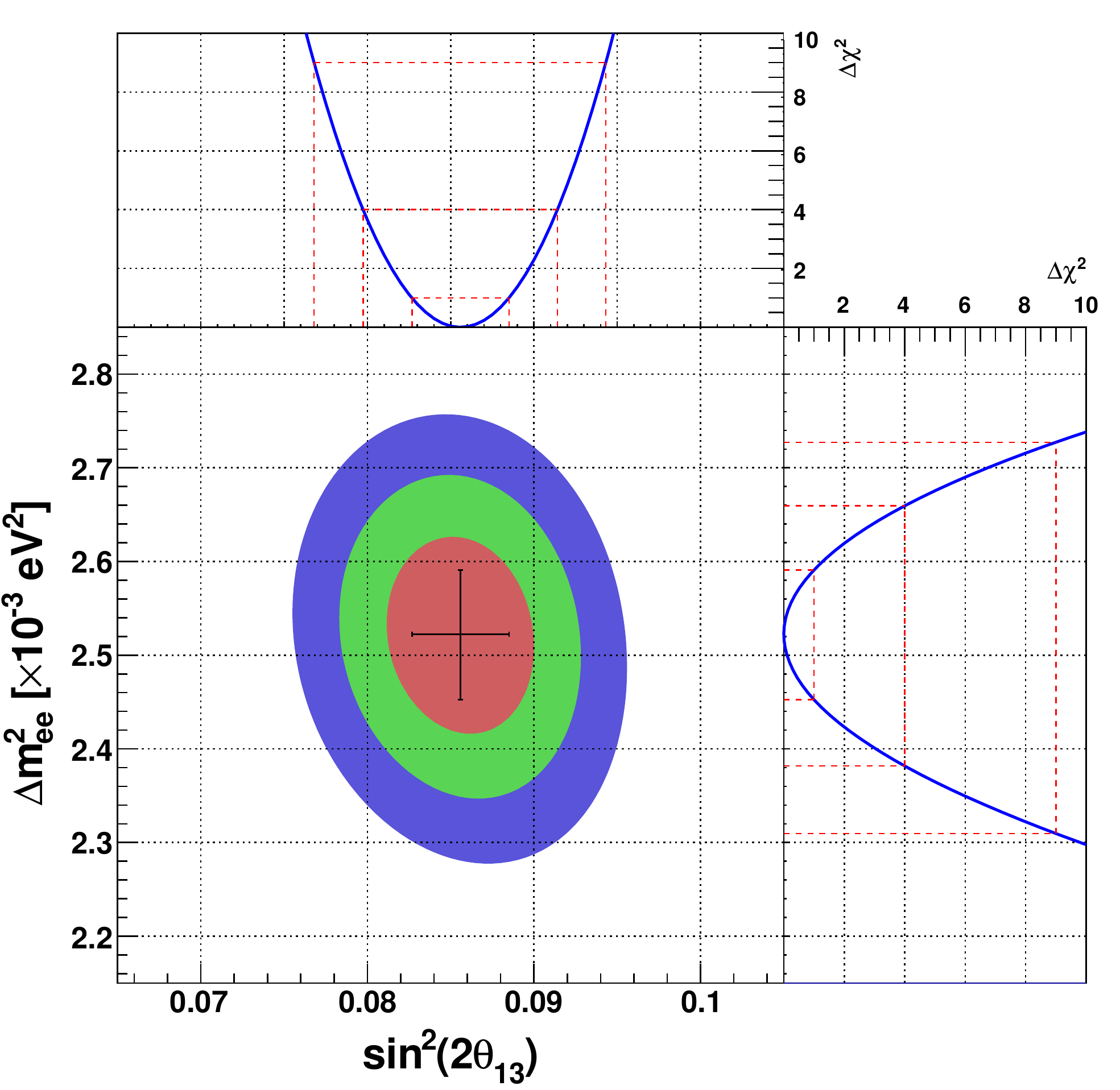}
\caption{The $68.3\%$, $95.5\%$ and $99.7\%$ C.L. allowed regions in the \dmee-\thet~plane.
		  The one-dimensional $\Delta \chi^2$
			for \thet~and \dmee~are shown in the top and right
			panels, respectively. The best-fit point and one-dimensional uncertainties are given by
      the black cross.}
\label{fig:contours}
\end{figure}

The reconstructed prompt energy spectrum observed in the far site is
shown in Fig.~\ref{fig:spectracomp}, as well as the best-fit predictions.
The $68.3\%$, $95.5\%$, and $99.7\%$ C.L. allowed regions in the \dmee-
\thet~ plane are shown in
Fig.~\ref{fig:contours}.

In summary, new measurements of \thet~and \dmee~are
obtained with 1958 days of data and reduced systematic uncertainties. This is the most precise measurement of \thet,
and the precision of $\Delta m^2_{32}$~is comparable to that
of the accelerator-based experiments~\cite{MINOS:dm2,NOvA2018,T2K2018}.

Daya Bay is supported in part by the Ministry of Science and
Technology of China, the U.S. Department of Energy, the Chinese
Academy of Sciences, the CAS Center for Excellence in Particle
Physics, the National Natural Science Foundation of China, the
Guangdong provincial government, the Shenzhen municipal government,
the China General Nuclear Power Group, Key Laboratory of Particle and
Radiation Imaging (Tsinghua University), the Ministry of Education,
Key Laboratory of Particle Physics and Particle Irradiation (Shandong
University), the Ministry of Education, Shanghai Laboratory for
Particle Physics and Cosmology, the Research Grants Council of the
Hong Kong Special Administrative Region of China, the University
Development Fund of The University of Hong Kong, the MOE program for
Research of Excellence at National Taiwan University, National
Chiao-Tung University, and NSC fund support from Taiwan, the
U.S. National Science Foundation, the Alfred~P.~Sloan Foundation, the
Ministry of Education, Youth, and Sports of the Czech Republic,
the Charles University Research Centre UNCE,
the Joint Institute of Nuclear Research in Dubna, Russia, the CNFC-RFBR
joint research program, the National Commission of Scientific and
Technological Research of Chile, and the Tsinghua University
Initiative Scientific Research Program. We acknowledge Yellow River
Engineering Consulting Co., Ltd., and China Railway 15th Bureau Group
Co., Ltd., for building the underground laboratory. We are grateful
for the ongoing cooperation from the China General Nuclear Power Group
and China Light and Power Company.

\bibliographystyle{apsrev4-1}
\bibliography{DYB_osc_2018}

\end{document}